# The Prosumer Economy
## Being Like a Forest

(v.1.2 - January 1, 2019)

by
Dr. Uygar Özesmi
Founder & Instigator, Good4Trust.org
uygar@good4trust.org +905336553689

**Abstract**:

Planetary life support systems are collapsing due to climate change and the biodiversity crisis. We have a very short time. Within 30-40 years, human civilisation may collapse if current trends continue. The root-cause is the existing consumer economy, coupled with profit maximisation based on ecological and social externalities.

Trends can be reversed, civilisation may be saved by transforming the profit-maximising consumer economy into an ecologically and socially just economy, which we call "the prosumer economy". **Prosumer economy is a macro-scale circular economy with minimum negative or positive ecological and social impact, an ecosystem of producers and prosumers, who have synergistic and circular relationships with deepened circular supply chains/networks, where leakage of wealth out of the system is minimised.** In a prosumer economy there is no waste, no lasting negative impacts on the ecology and no social exploitation. The prosumer economy is like a lake or a forest, an economic ecosystem that is productive and supportive of the planet.

We are already planting this forest through Good4Trust.org, started in Turkey. Good4Trust is a community platform bringing together ecologically and socially just producers and prosumers. Prosumers come together around a basic ethical tenet "the golden rule" and share on the platform their good deeds. These good deeds may include charity, volunteering, goods and services they bought on the Good4Trust online bazaar to support ecologically and socially just production. The relationship are already deepening and circularity is forming to create a prosumer economy. The platform's software to structure the economy is open-source, and is available to be licenced to start Good4Trust anywhere on the planet.

Complexity theory tells us that if enough agents in a given system adopt simple rules which they all follow, the system may shift. The shift from a consumer economy to a prosumer economy has already started, the future is either ecologically and socially just or bust.

### 1 Introduction

#### 1.1 The looming collapse of civilisation

The etymology of the word "ecology" comes from "oikos", the house and its garden in ancient Greek, it is the science of our house, the Earth. Scientists write that since 1980 the Earth has surpassed its capacity to renew itself. Maybe our biggest problem is climate change, which could cause dangerous and irreversible ecological and social catastrophe if the global average temperature increases more than 2°C (IPCC, 2018).

The carbon dioxide level in the atmosphere before industrialisation was 280 parts per million (ppm). In May 2018, at the Mauna Loa Station of the US National Oceanic and Atmospheric Administration in Hawaii, this concentration was measured to be 411 ppm. The fact that the level of the carbon dioxide in the atmosphere has surpassed 350 ppm shows that we have exceeded the safe limit. This 350 ppm limit was broken for the first time in January 1988. The level of carbon dioxide concentration before the industrialization era had not surpassed 300 ppm for 800 thousand years. By crossing the 400 ppm carbon dioxide concentration we have increased the average temperature by 1.1°C. (IPCC 2015).

An average increase of 1°C may not seem to be much, however 1°C corresponds to an increase of 7 to 15°C in extreme weather conditions, and scientist have shown widespread persistent changes in temperature extremes in the 21st century, earlier than predicted before (Li et al, 2018). To illustrate this I will use an example from Turkey. Turkish State Meteorology Agency (Meteoroloji Genel Müdürlüğü, 2016) data show that between 15-19 February 2016 in Milas Muğla the temperature was 32.4°C, whereas before the highest ever recorded February temperature extreme was 24.9°C in 1950. In other words, there was a difference of 7.5°C in extremes.

Let's expand a little from Turkey, and look at 2018 extreme temperature records around the world. In 2018, the highest recorded temperature in London was 36.7°C, 39.5°C in Istanbul, 39.7°C in Paris, 40.0°C in New York, 41.1°C in Japan, 48.9°C in Los Angeles. Let's assume they all contain the average 1.1°C temperature shift (more than 7°C in the extremes) just like in Milas, Muğla, Turkey, already. If we don't change our way of consumption and use of products such as coal, oil, natural gas and other fossil fuels then the temperature would increase 2,6-3,2°C by 2050 (Brown and Caldeira 2017, 47). So we are foreseeing about 15°C additional in extremes… So let's add 15°C to see what extreme temperatures we would expect in 2050: London 51.7°C, Istanbul 54.5°C, Paris 54.7°C, New York 55.0°C, Japan 56.1°C, and 63,9°C in Los Angeles. In a specific research and under a conservative scenario Bador *et al* (2017) estimate temperatures exceeding 50°C in France in 2100. Probably by 2050 or cautiously 2100, these temperature would be deadly for humanity (Mora et al, 2017). Global warming is happening at an ever increasing rate (Xu et al, 2018).

To avert the collapse of civilisation, the success of The United Nations Framework Convention of Climate Change (UNFCCC) is critical but progress is slow, and the threat is fast and monstrous. Unfortunately, governments and political leaders are still not facing this huge issue. Climate change brings storms and floods, as well as droughts causing a fresh water crisis. Climate change leads to agricultural collapse and hygiene problems. Humanity would face widespread famine and disease, at extremes, roads and cables may melt and we might have a total collapse of civilisation. Quite a dystopia…Before we move onto the solution, we need to clarify another great threat.

Biological diversity, including humans, is the diversity of living beings and their habitat, it is defined by the diversity of relations between each other and their environment. This diversity is disappearing today. According to the report of WWF in 2016, populations of vertebrate animals—such as mammals, birds, and fish—have declined by 58% between 1970 and 2012. And we're seeing the largest drop in freshwater species: on average, there's been a catastrophic 81% decline in that time period. The terrestrial Living Planet Index (LPI measures population sizes of vertebrate species) shows that populations of terrestrial species have declined by thirty-eight percent overall between 1970 and 2012. The freshwater LPI shows that on average the abundance of freshwater species populations monitored in the freshwater system has declined by eighty-one percent between 1970 and 2012, The marine LPI shows a thirty-six percent overall decline between 1970 and 2012 (WWF, 2016).

Currently, out of 149 million km$^2$ total land area of the planet, including deserts and glaciers, we have transformed 51 million km$^2$ into agriculture. Of the habitable area 50% is used for humans, apart from glaciers and deserts (Figure 1). Out of the agricultural areas 77% is used to grow feed for animals which we have domesticated. With the remaining 23% we feed ourselves.This 77% agricultural land allocated to animal feed only constitutes 17% of human calorie intake and 33% of protein intake. While the rest of the 83% of calories and 67% of protein needed for human consumption are produced on 23% of the available agricultural land. Livestock takes up nearly 80% of global agricultural land, yet produces less than 20% of the world's supply of calories (Ritchie 2017).

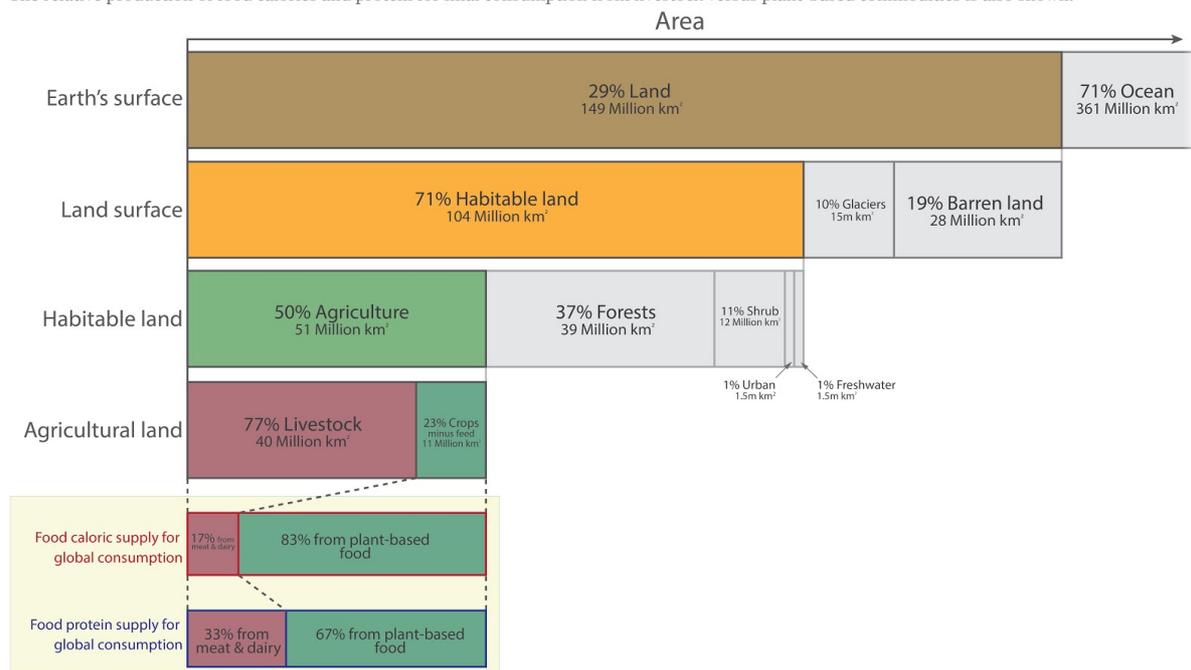

Figure 1. Global surface allocation for human use (Richie and Roser, 2017)

With this level of terrestrial human encroachment we have destroyed habitat for other large living animals on land. The result is incomprehensible, wild terrestrial vertebrates are at only 3% in mass as compared to 'man and his animals' which are 97% of the total mass (MacCready, 2004). We have invaded Earth, even before the Martians. Marine ecosystems and fish populations are also suffering drastic damage, and declines threaten many marine animals. Data from more than 230 fish populations reveal that there is a median reduction of 83% in their breeding population size from known historic levels (Hutchinson and Reynolds 2004). Scientists estimate we're now losing species at 1,000 to 10,000 times the background rate, with literally dozens going extinct every day (Chivian and Bernstein 2008). It could be a scary future indeed, with as many as 30 to 50% of all species possibly heading toward extinction by mid-century (Thomas *et al*, 2004).

If we want to stay alive, if we believe in the rights of other species to exist and the rights of Mother Earth, our impact on the planet has to be reduced. This is not an issue concerning a country, an ethnic group, a religion or even humans only, it is our problem, all living beings. The root cause of this devastation, this debacle, this collapse is an economy that is based on exploitation of nature and other fellow humans, an economy based on competition, growth, consumption and profit-maximisation. Thus, we must seek the root cause of the environmental crisis in the dominant socio-economic paradigm we live in (Figure 2). Professor Guy McPherson from the School of Natural Resources at University of Arizona puts it right in our face: "If you think the economy is more important than the environment, try holding your breath while counting your money."

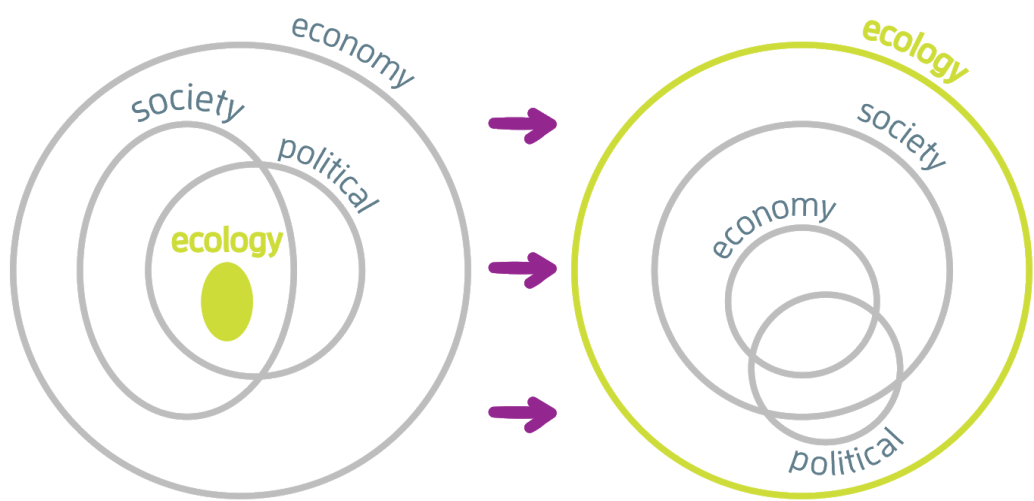

Figure 2. Shifting the dominant socio-economic paradigm.

Without shifting the dominant socio-economic paradigm that is based on competition, growth, consumption, and profit maximisation, we will not be working for a sustainable future. That sustainable future we seek, must move beyond the economistic definition revolving around "development that meets the needs of the present without compromising the ability of future generations to meet their own needs." Intergenerational equity is fair, but considering both economy and ecology share the same root "oikos" we need to frame the world not by economy but by ecology… The law needs to follow the science, so in this new oikos-centric paradigm ecology hosts society, and their politics. Ecology sets the limits in which economy and politics govern society. In that sense the new paradigm and what we seek is transformationist (Hopwood 2005) and works for a great transition (Raskin 2016).

**1.2 We can save civilisation through a different economic system**

Today when we spend 1 USD on average in a low-income country we emit 1.011kg of carbon dioxide equivalent, whereas 0.583kg in a high-income country (Caron and Fally, 2018). Essentially as humanity, we destroy the environment with every dollar we spend. When we decide to be part of the killing of an animal and purchase 1 kg of meat we also emit 28.7 kg of carbon dioxide equivalent, and produce 4 kg of waste, when we buy a pair of trousers we emit 6.3 kg carbon dioxide equivalent, and produce 25 kg of waste, every time we renew our smart phone because of planned obsolescence we produce 110 kg carbon dioxide equivalent, and 86 kg of waste not regarding the use (Laurenti, 2016). 1 kg of industrially produced meat has an unbelievable water footprint of 15500 litres (Hoekstra 2008), 1 pair of cotton trousers 3233 to 4894 lt (Chico et al 2013) and a smartphone requires 18$m^2$ of land and 13000 litres of water (Burley 2015, also see [watercalculator.org](watercalculator.org)). Due to our economic activity, we have passed the yearly regenerative capacity of the Earth since the 1980s and in 2018 World Overshoot day in terms of our global footprint has been August 1, according to [footprintnetwork.org](footprintnetwork.org).

In terms of the human sphere, things are also in disarray. Repeated economic turmoil such as the subprime crisis, the fiscal crisis in Turkey in 2018, or any other crisis at national, regional or global level, are symptoms of economic unsustainability. These crises indicate the need for a new economic system. Even if we were able to implement technological fixes such as producing all our energy from photovoltaic solar panels in the existing consumer economy, then emissions from other sectors such as agriculture, as well as the biodiversity crisis would still continue, and we would continue to cause exploitation of humans around the planet.

Similarly primarily as a technological fix with social and economic dimensions, the circular economy, and green growth is also proposed as a solution. But they are criticised for being primarily focused on large-scale technological solutions, caring little about politics, human rights, not recognizing social actors, and that the world as we know it can continue with green growth (Fatheuer, 2016). For example, today the electric car and electronics battery sector is implicated in more than 10,000 children worked in abhorring conditions in the cobalt mines of the Democratic Republic of Congo (Walt and Meyer, 2018).

Our current global economy is obviously not inclusive. There is no just distribution of wealth, instead it is concentrated in the hands of a few. The economy is based on massive production, competition, growth, consumption and profit maximization creating dissonance. In order to lower the prices of products, and to sell more, social dumping, and slave or child labor are seen as acceptable. People work in inhumane conditions, furthermore they are losing their place to machines in this madly competitive market.

**The current dominant socio-economic paradigm based on competition, consumption, and profit-maximization through externalities is the main cause of climate change and the biodiversity crisis facing our planet.** This system is indexed to an illogical expansion, to growth, however resources are limited, and exploitation of other beings can only go so far. Scientists agree that the current level of exploitation will lead to irreversible damages to our life support systems and even be the end of human civilization.

**Therefore we need a new economic system that is based on ecology to meet the needs of the planet, and social justice to meet the needs of people.** In an inclusive economy, the production of goods and services are optimized to increase the prospects and opportunities for everyone. Competition is not seen as a must, but as an ill of society. Mutualism, and reinforcing, supportive cycles are sought after, win-win scenarios are created, in case of losses, they are compensated and repurposed. Nature is not seen as a resource but as a being. In order to prevent the reduction in structural biodiversity, and collapse of civilisation, **we propose a new economic system that we call the Prosumer Economy.**

### 1.3 The Prosumer Economy

For the purposes of Prosumer Economy, a prosumer is a person who treats others as they would like to be treated themselves (golden rule) hence creates value with their actions for society and the planet. They support social and ecologically just productions directly. The "other" to the prosumer is all living and non-living beings on this planet. Hence the golden rule necessitates a harmonious existence with the planet as an element of it. To quote Defne Koryürek, one of Good4Trust's Council of Seven members in Turkey: "Prosumers are the ones who refuse the two-polar definition of the growth economy knowing that every producer is also a consumer and every consumer is a producer." This blurring of the roles of consumers and producers is said to originate in the cooperative self-help movements during economic crises going back all the way to the Great Depression starting in 1929.

Marginally relevant to our topic, the specific term was first defined in 1980 (Toffler, 1980) to describe a proactive consumer who is involved in the design and improvement of goods and services. In our case, the prosumer is the reason for the good and service. But since 1980 the term has evolved to describe a consumer who consumes what they produce, and this in multiple forms, which may or may not involve brands or the market (Paltrinieri and Esposti, 2013). An example would be a person who produces the electricity they use on their roof, through solar photovoltaic panels. The closest we come to what prosumer in the prosumer economy means, is defined by Paltrinieri and Esposti (2013) as "a commons based prosumerism may arise from the use of social media to promote civic engagement, and to build relationships based on specific interests and mutual trust, through a process in which citizenship is activated in actions of mutual support." As Alhashem (2016) has observed prosumption is a discursive practice of consumer empowerment through self-discipline and collective education in contrast to other exploitive practices such as consumer co-creation by corporations. Our approach is parallel with activists who embrace prosumerism as a way to exclude and resist the profit-maximising corporate producer. Although supporting other small and local producers may seem irrelevant to people with diminished disposable income caused by various economic trends such as globalization, automation, and wealth concentration, prosumer economy at the local level means local development and solidarity, as everyone becomes a producer to make a living, hence a prosumer.

I define the **Prosumer Economy as a macro-scale circular economy with minimum negative or positive ecological and social impact, an ecosystem of producers and prosumers, who have synergistic and circular relationships with deepened circular supply chains/networks, where leakage of wealth out of the system is minimised.** Prosumer economy is like a forest, productive and beneficial to our Planet.

The exodus from consumption, and the building of the prosumer economy will need a transition process, from a mechanical economy that thinks it can get bigger and faster to one that is organic and homeostatic. In this shifting conceptualisation for a prosumer economy we believe that civil society, and grass roots are critical in influencing and shaping political and economic decisions. If enough number of people, and especially young people organise, and believe in real and solid change, and then act on it the transition is possible. For this ordinary people need to gain power, have a strong voice, connect with people who share their concerns, and act together for social change. This grassroots conceptualisation and organisational language is immediately attributed to left movements and is labelled as socialist, or even as communist as a reaction. Please note that the same language is also used by the right movements. These bottom-up approaches and collective action is not reserved to any ideology, but are tools of social change. Prosumer Economy refuses to succumb to sides or frames and categorization of the past, and has a meta-vision (Özesmi, 1999). Old governance systems are failing to understand that their realm does not exist anymore, we need to move beyond our known paradigm, and restore our environment for the future that now belongs to young people, and for this the economy needs to change.

## 2 "Young" people demand a different market

In developed countries, the market for organic, locally produced goods and renewables is growing strong, and steady. Local food sales in the U.S.A. grew from $5 billion to $12 billion between 2008 and 2014, and it is predicted that local food sales would jump to $20 billion in 2019, outpacing the growth of the country's total food and beverage sales. Also the 2013-14 school year saw nearly $790 million worth of local foods purchased by primary and secondary schools, and by 2015, nearly 8,500 farmers markets had sprung up across the U.S.A., an increase of more than 380 percent in 20 years (Hesterman & Horan, 2017).

According to 2010 statistics there were 37 million hectares of agricultural land in organic production, managed by 1.6 million certified producers. Non certified producers, for which no data exists, are estimated to be many times this figure (Dittrich, 2012). In 2016 the total certified land in organic agriculture reached 57.8 million hectares. In 2003 25 billion, in 2008 50.9, and in 2016 90 billion US dollars worth of organic products were sold globally (Willer & Lernoud, 2018). The demand and supply is steadily increasing.

A consciousness is developing about products that damage the environment, human health and the producer itself. There is a wave of conscious consumerism, albeit still an individualistic act. People are trying to buy local, fair-trade and organic products, for example in Germany 84% of people regardless of age group would like to know whether the product was fair, 82% whether it is produced environmentally friendly, 81% without Genetically Modified Organisms, and 79% would like a government enforced animal welfare label (BMEL, 2017).

Not only in foods, but also in the energy sector there is a rapid transformation. European Union has a target to cover 20% of its energy needs with renewable energy resources by 2020, and then 50% by 2050. As Ntanos et al. (2018) found that renewable energy investment is correlated with increased GDP growth and labor force in Europe. Annual investment in renewable power and fuels was 274 billion USD in 2016, and 279.8 billion USD in 2017 globally. Solar power as photovoltaics only has grown 30% in the last year to 402 GW (REN21, 2018).

The shift towards ecologically conscious consumer behavior is also not limited to the West or North. Indian consumers are increasingly becoming ecologically concerned in their buying behavior, research examining their pro-environmental consumer behavior is found in several studies (Taufique and Vaithianathan 2018).

There is a return to what is deemed natural. Yet living by these ecological and social principles is still complex in terms of finding and trusting products and producers. The demand side "conscious consumers" and the supply side "conscious producers" remain largely unorganized. They accept the competition in the market as a norm, and cannot think outside of the "competition is good for us" paradigm. By consequence most of the people feel helpless when it comes to face the environmental issues. However, new generations currently between 19 and 35 years of age - accounting for 27% of the global population or about 2 billion people, are fast becoming the world's most important generational cohort for consumer spending growth, sourcing of employees, and overall economic prospects. They are different from previous generations in driving a collaborative economy, leading the way in terms of embracing (or at least experimenting with) new formats such as crowdfunding and peer-to-peer transactions. Surveys show that 36% of Millennials are willing to buy eco-friendly products (Rogers 2013) and demand a more environmentally sensitive workplace (Guevarra 2010). The young generation chooses significantly more ecological products, are more active than other groups on environmental issues, because they have ability to use borderless technology to communicate and exchange information. The social network of young people has a tendency to dominate their perception, and this reinforces their personal affective response and refines their ecological knowledge and have significant more green intentions (Kanchanapibul *et al*, 2014).

Change is happening and it is driven by people and especially young people who want to **"consume" in a way that is better for the planet and the health of their communities. There is a demand for a new economy in the grassroots.**

### 3 The Prosumer Economy as a new grassroots economic governance

Prosumer economy is inclusive by constitution, in other words system design. The system circulates wealth among small scale ecologically and socially fair producers. Prosumer economy creates a circular economy, not only at the product or service level, but more importantly at the macro-level, among producers. The wealth generated starts going from one hand to the other and creates true economic and social development through circularity. Luckily, grassroots movements are taking shape that can easily form strong synergies and bonds with the prosumer economy.

The degrowth movement (https://www.degrowth.info) has similar aspirations to the prosumer economy. Degrowth defines itself as striving for the good life for all, which includes deceleration, time welfare and conviviality. "A reduction of production and consumption in the global North and liberation from the one-sided Western paradigm

2018 U. Özesmi - Prosumer Economy    10/10

of development. This could allow for a self-determined path of social organization in the global South. An extension of democratic decision-making to allow for real political participation. Social changes and an orientation towards sufficiency instead of purely technological changes and improvements in efficiency in order to solve ecological problems." The degrowth movement believes that decoupling economic growth from resource use is not possible, and open, connected and localized economies, need to be built. In this sense the prosumer economy could be considered an associate to the essence of the degrowth movement.

P2P movement also shares many similarities with prosumer economy. P2P is an abbreviation of "peer to peer", "person to person" or "people to people". It includes the creation of common goods through open, participatory production and governance processes to which there is universal access through licenses such as Creative Commons, GPL, Peer Production Licence. The similarity becomes more pronounced in its streams, where P2P Cultures and Politics is designed to help build a relatable identity and culture for the Commons, featuring inclusivity, gender equality and diversity, based on these values aims to transform progressive politics in a commons way, inside and outside the institutions. Open Coops & Sustainable Livelihoods guides P2P in examination of labor, carework, well-being and emancipation for commoners, and the creation of durable, transnational networks to construct ethical markets. They work on creating an Open Source Circular Economy to create synergies between cooperative peer production and sustainability. To do this, they attempt to show how a transition to new modes of production, governance and ownership can solve ecological and climate crises (https://p2pfoundation.net). In this sense the Prosumer Economy has a similar if not the exact goal.

Another grassroots movement is the transition network of communities coming together to reimagine and rebuild the world (https://transitionnetwork.org). The communities are reclaiming the economy, sparking entrepreneurship, reimagining work, reskilling themselves and weaving webs of connection and support. They also have a REconomy line of work in exploring the potential of community-led economic change.

But maybe the oldest and most common grassroots movement are cooperatives. Cooperatives are an essential part of prosumer economy. The 7 principles of cooperatives are also ones that prosumer economy values; (1) voluntary, open ownership, (2) democratic owner control, (3) owner economic participation, (4) autonomy & independence, (5) education, training & information, (6) cooperation among cooperatives, (7) concern for community. But in a much broader framework we share the overarching aim of "The Social Solidarity Economy" as an alternative to capitalism and other authoritarian, state- dominated economic systems (http://www.ripess.org). It is "ordinary people that play an active role in shaping all of the dimensions of human life: economic, social, cultural, political, and environmental. Social Solidarity Economy exists in all sectors of the economy production, finance, distribution, exchange, consumption and governance. It also aims to transform the social and economic system that includes public, private and third sectors… Social Solidarity Economy seeks systemic transformation that goes beyond superficial change in which the root oppressive structures and fundamental issues remain intact." Prosumer Economy is in alignment with all major grassroot movements, and for surely more in the existing diversity and multiplicity, and hence hopes to create synergies and integration wherever possible.

The governance of prosumer economy is as the name indicates in the hands of prosumers, who source their needs from ecologically and socially just goods and services from producers. Everyday, they vote, when they take out their wallets and invest in a product or service. **As producers also source their inputs from similar producers, they deepen the supply chain in the prosumer economy. Everyone becomes a prosumer.** The success and well-being of these producers, will pull popular opinion to support the sustainable alternatives, as such producers in the consumer economy will also be pulled in. This will push the unsustainable alternatives, the ecologically and socially destructive profit-maximising, and externalising ones from the center to the margins. Power will shift with the economy.

Prosumers who are part of the modes of production will know who is producing the products and services. Knowing the producers, they will trust that just working conditions are ensured, the impacts on the environment are minimized, and every one's health is cared for. The prosumer can make sure and check practices as the system is based on full transparency, and accountability. Similar systems of ethics, trust and standards have emerged from Anatolia in the past such as "Ahilik" (Kantarcı 2014). *Ahi* organisations were present in all craftsmanship and trade businesses and were ensuring that there was no competition but, honesty, justice, customer satisfaction, and highest quality of production. They governed the bazaar of all cities, small and large. In the prosumer economy similarly we bring together ecologically and socially just producers and prosumers in shaping an economy based on ethics. As in a bazaar producers are preferably small scale and local.

As in Ahi organisations, still today in line with tradition, the Grand Bazaar in Istanbul is governed by the shop owners, but this time thru an association formed in 1952 in a legalized "nation state" environment. One could argue nation state democratic processes, or multilateral dialogue and agreements that lead to pro-environmental and prosocial legislation would be helpful in creating the aspirations of the prosumer economy. I might be sympathetic to such an idea and am happy to see it when it happens. Unfortunately, they have not worked in full, the situation we are in today speaks for itself. Although it could work under different governance regimes, today I am more concerned that these systems of legislation, policies, and incentive structures governed and implemented thru the state mechanisms are perverse, such as subsidies that go into the fossil fuel industry or conventional agriculture (Myers and Kent, 2001). Hence I want to go back to the attitude of a philosopher from Anatolia, Diogenes of Sinope, when it comes to state matters. Alexander the Great found Diogenes lying in the sun and asked him if he would wish anything from him. "Yes," said Diogenes, "stand a little out of my sun."

## 4 The Transition to Prosumer Economy

Let us summarize how the transition to a new prosumer economy will happen and transform the system. As was made clear in the beginning, currently our economy is based on the exploitation of humans and nature. In order to grow the economy, the competitive system asks for profit-maximisation, and as it grows in general it becomes more profitable with a positive feedback cycle. Maximising profits, leads to externalisation costs, and growth leads even to more externalities towards human and nature. Even the applauded new disruptive "sharing" economies are externalising real estate, vehicle and other traditional costs to the individuals, the person in the street. Imagine where and how a mining industry, a coal power plant, a chemicals, or an arms production facility externalizes costs. In the consumer goods market planned obsolescence is a pandemic, that only encourages the *buy and throw away* culture. Unfortunately most circular economy efficiencies are gained to reduce costs, rather than to minimise the ecological and/or social damage.

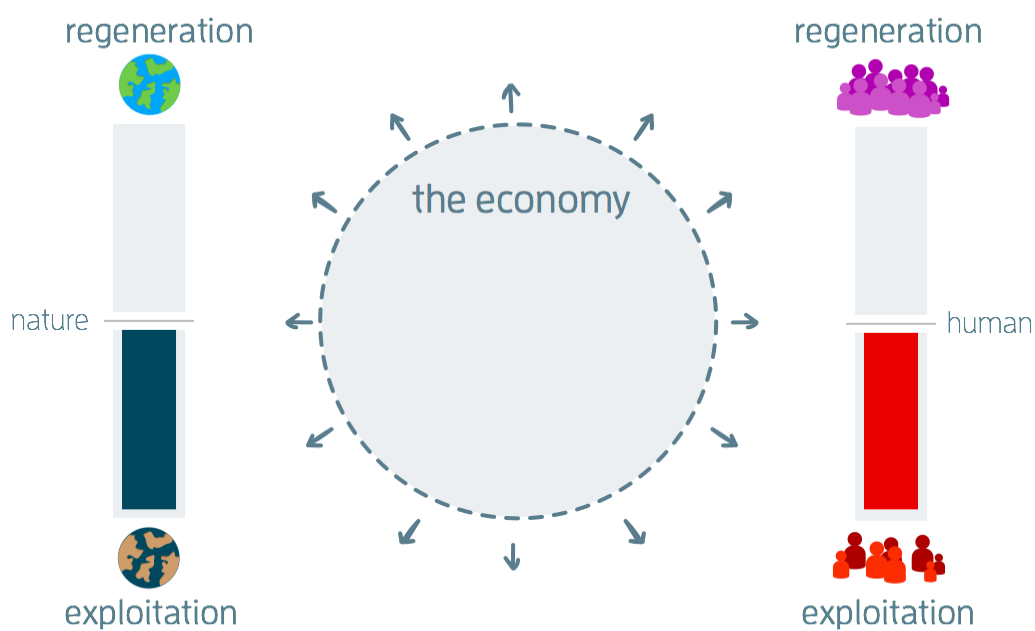

Figure 3. The growth of the economy that is exploiting the planet and people. The economy has reached its limits, where planetary life support systems are collapsing.

As seen in Figure 3, the economy in the current socio-economic paradigm is a balloon that will eventually explode. It is based on the myth of the perpetual motion machine. Infinite growth in a finite world is simply against physics. In reality it is a dirty and oily internal combustion engine, burning fossil fuels, spewing pollution, making human and nature deadly ill. The consequences are climate change and the biodiversity crisis. Technological optimists may want us to believe that all these problems can be solved by human ingenuity, by science and technology. We have already proven that reliance on such hubris and human arrogance has brought us more problems. With plastics as a great solution to our problems, we have created giant microplastic islands in the West and East Pacific Ocean, a 19 by 5 km plastic island swimming in the Caribbean Sea, whales that die with 29 kg of plastic in their stomach, albatrosses in remote islands, and mammals in deep forests dying of the plastic they ingest, finally microplastics in human urine. There are countless more like Chernobyl, Fukushima, DDT and other persistent organic pollutants, and lately the neonicotinoid and glyphosate crises. However, we don't need to succumb to our hubris. There is a way out of the climate and biodiversity crises. We can have a strong and vibrant economy to fulfill our needs and earn a livelihood.

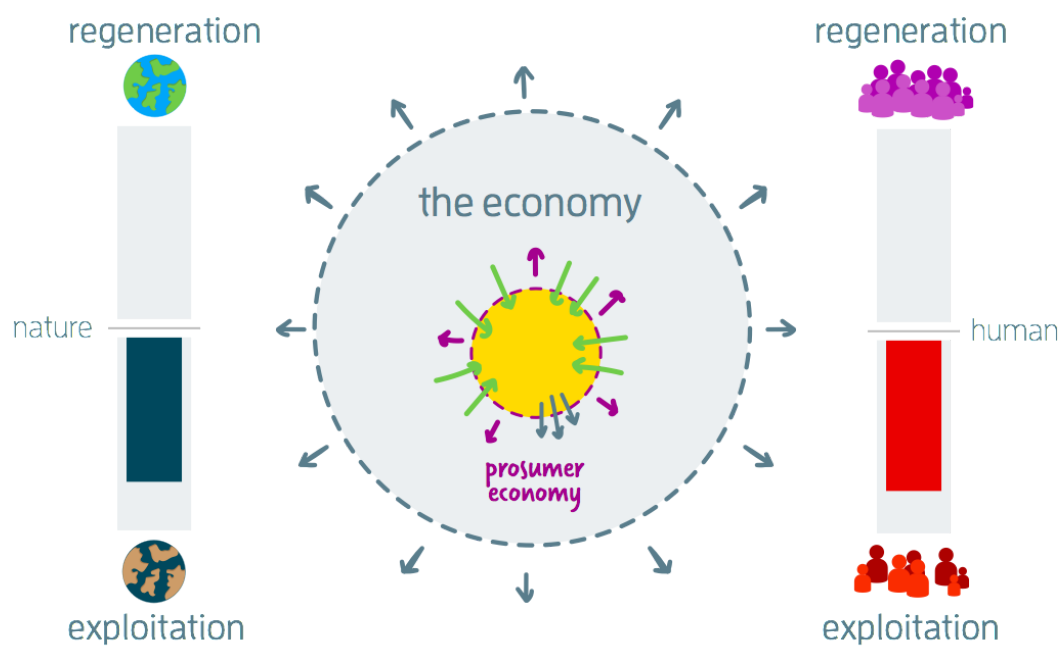

Figure 4. A prosumer economy seeded into the existing economy, starts eating consumption, as leakage of wealth and materials to the outside of prosumer economy is reduced, the economy will slow down in growth.

As seen in Figure 4 let's seed this consumer economy with a prosumer economy. The prosumer economy inside this system, will grow into it, if the wealth that goes in is more than the wealth that goes out. Essentially we are building towards a non-leaking system. This will reduce the overall growth and hence the negative impact on people and the planet. One may visualize a bacterial growth in a petri-dish to get a sense of the transformation process. As the bacteria grow in the petri-dish by eating the agar, the food is essentially transformed into more bacteria Figure 5. Consider the consumer economy the food of the prosumer economy.

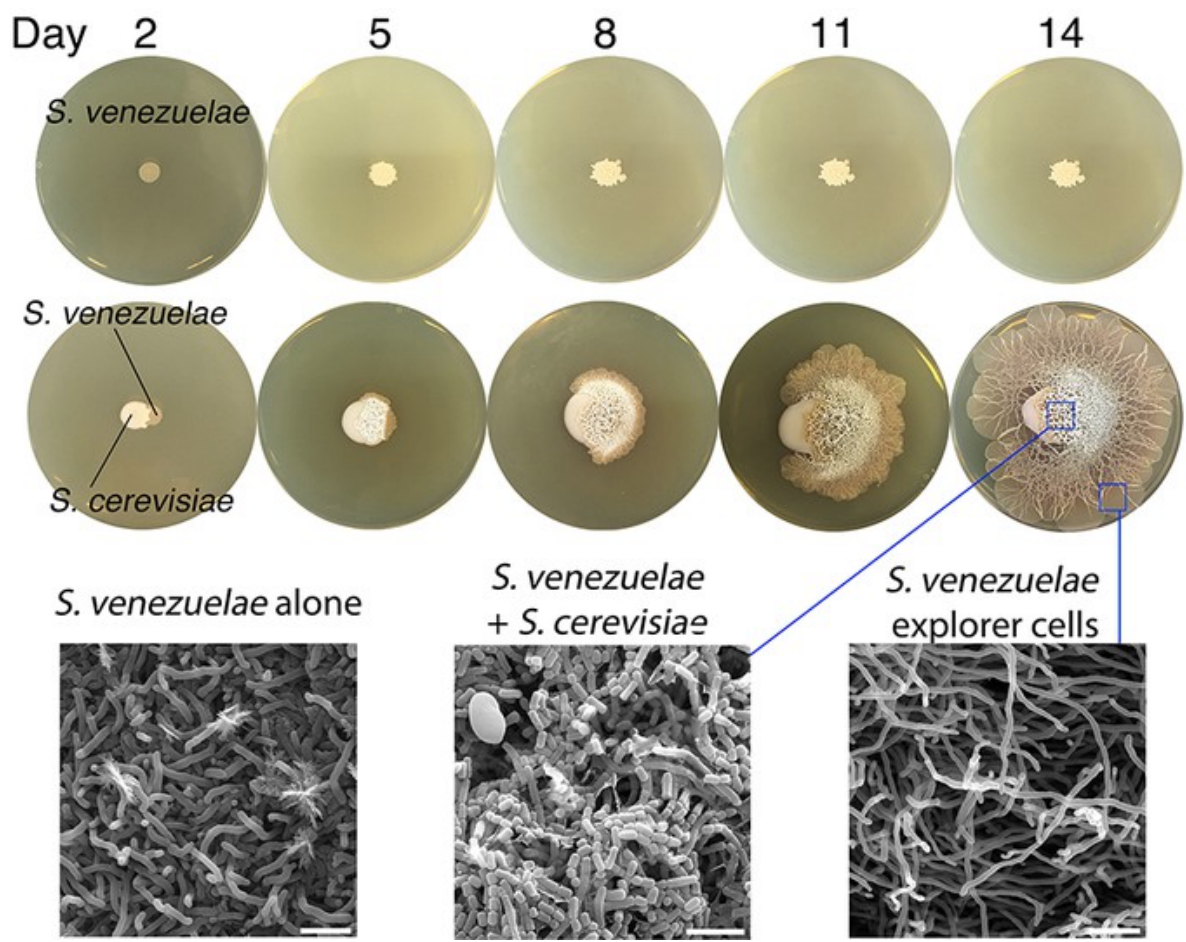

Figure 5. Synergistic bacterial and fungal growth in a petri-dish over time, transforming the agar into bacteria and fungus. Please note that synergistic

cooperation between two kingdom leads to elevated levels of transformation (Jones et al. 2017).

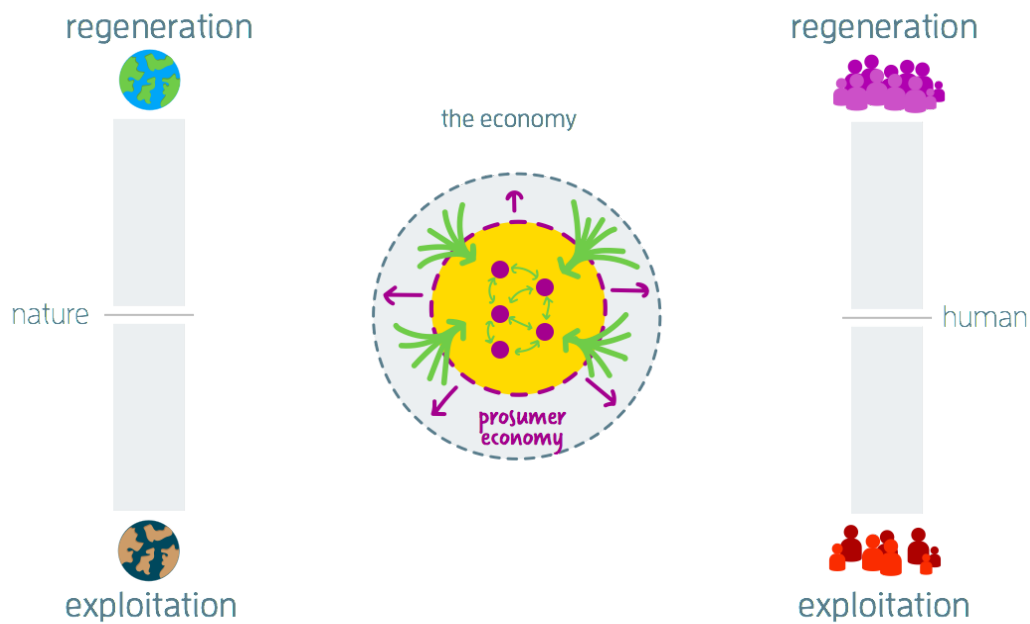

Figure 6. As the leakage is eliminated and synergistic macro-scale circular economic relations are established, the consumer economy will be consumed and will stop expanding and start shrinking.

If we eliminate the leakage completely, and start building a macro-level circular economy, that does not produce waste and exploitation, in other words eliminates externalities, not only would we be able to stop the growth of the consumer economy, but we would be able to stop the destruction of nature and exploitation of people (Figure 6). The macro level economy similar to the mutualism between the bacteria and the fungi (Figure 5) is based on synergistic win-win constructions. Producers in the system will provide for each other, they will inspire and find purpose together. It is an ironic myth of the current socio-economic paradigm that competition makes things "better", more "efficient" and leads to "progress" besides those statement being all dependent on the "for whom" context, it simply does not hold true empirically, the opposite is true compared to cooperation (Kohn, 1992). Competition is not an inevitable part of "Nature" and definitely not of "Human Nature". In Nature, which is not an ethical role model for human society, it barely exists, and is less effective than cooperation in natural selection (Kropotkin, 1902).

### 4.1 Regeneration after transition

Once the prosumer economy would be "the economic system", essentially having transformed to such a level that it encompasses the whole system, we would create an economy that is not only circular but also regenerative. This transformation would satisfy needs and create happiness in the human condition and will create harmony with nature. The regenerative economy will be increasing synergies, rehabilitating and giving back to nature (Figure 7).

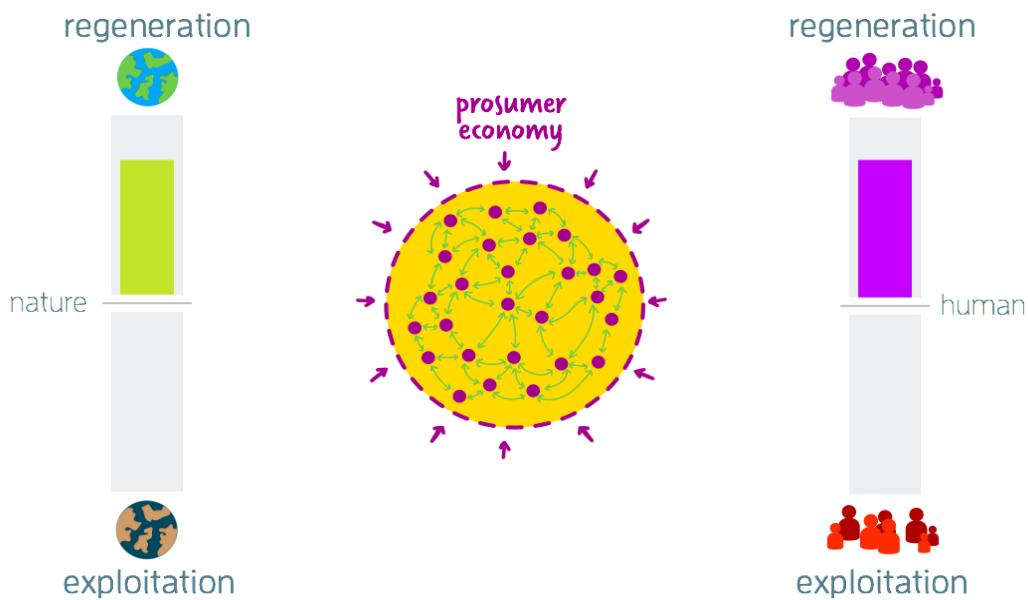

Figure 7. The prosumer economy encompasses the whole economy, and through synergistic interactions allows for regeneration of human relations as well as enriching biodiversity and planetary life support systems.

### 4.2 Being like a forest

In the final stage of the prosumer economy imagine a forest, a meadow ecosystem, or if you like something like the Baikal Lake. They are limited by physical boundaries, whether geomorphology, altitude, or climate. These are very complex ecosystems and they are very productive. They are full with life, containing a myriad of producers and consumers, or rather prosumers. The human economy is not necessarily more complex or more productive than these ecosystem are, yet while we destroy Mother Earth, they create it. So our challenge is to be like a forest. The prosumer economy is human economy in the structure of a forest (Figure 8).

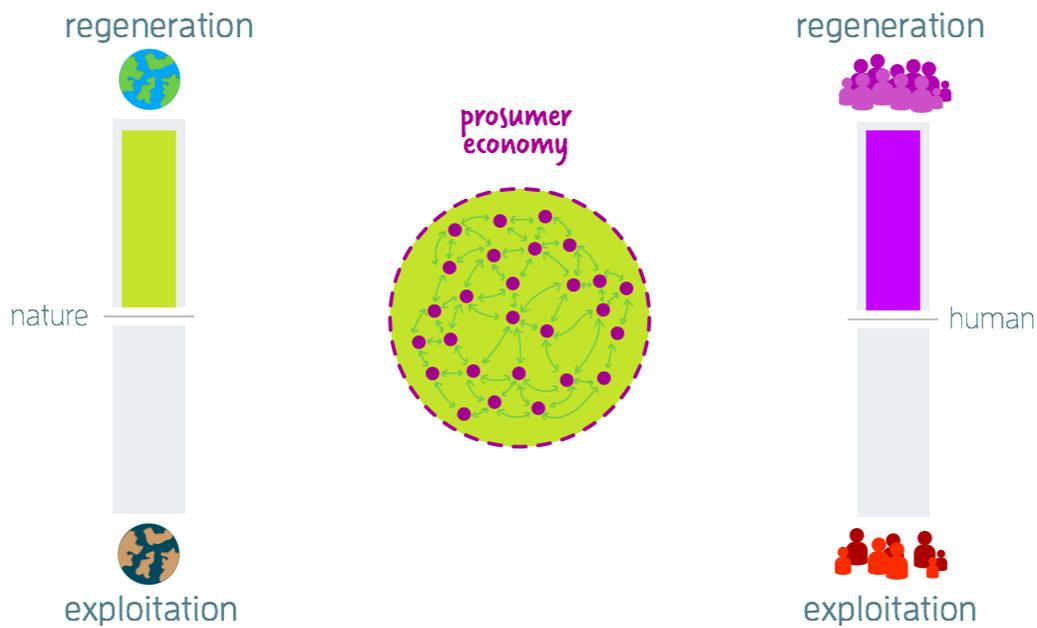

Figure 8. The prosumer economy as a forest, bound by physical planetary boundaries of what the sun gives us, and what space other beings occupy, a macro scale circular economy, where there is no waste and negative impact on planetary systems, where people exist in harmony with the planet as an additional human ecosystem.

Becoming a forest can only be accomplished one tree at a time, by constructing or stitching relations with patience and determination. A framework is needed for building such an economy. Therefore we have developed a platform called Good4Trust.org. In Good4Trust.org we plant tree after tree and keep adding squirrels, and birds, and ants, and bugs to grow slowly but surely this economic ecosystem. In a more mechanical definition Good4Trust is a disruptive internet technology platform acting as a utility to organize the prosumers into an economy. In the online platform, **pro**ducers and con**sumers** join hands to become **prosumers** and **build a prosumer economy**, a new circular economy formed and governed by the people participating in that economy.

## 5 Good4Trust: Prosumer economy in action

The prosumer is someone who is actively involved with the design, production and delivery of the goods and services they consume. Anyone can become a prosumer, anyone can register and participate. We assume that people joining the utility, the Good4Trust platform, want to lead a lifestyle based on values of goodness. Goodness is defined by the golden rule also written at the entrance of the United Nations, "Do unto others as you would have them do unto you." For Good4Trust "others" includes all beings on Earth. Good4Trust encourages pro-environmental and pro-social behavior. The platform has also a social solidarity aspect, prosumers are encouraged to write on the Good4Trust 'stream' their deeds or any behaviour they see that follows the golden rule. For instance, buying products and services that minimize harm to nature and society to meet their basic needs is considered the highest form of goodness or simple good deeds like contributing to the education of a child (Figure 9).

Figure 9. Good deeds "stream" on Good4Trust.org

2018 U. Özesmi - Prosumer Economy    19/19

Scientific research shows that doing good deeds is good for health (Whillans et al 2016), can be learned (Weng et al 2013), and makes you feel good and gives you a warm glow, a.k.a. helper's high (Andreoni 1990). Therefore the intrinsic motivational aspect of pro-environmental behavior becomes critical for a sustainable future (Van der Linden, 2015). Studies of brain chemistry and social behaviour supports and claims that seeing a good deed of a person stimulates the desire of also doing good deeds and this behaviour would spread and increase in frequency (Schnall et al. 2010).

Good4Trust structures a truly ecologically and socially just economy, enabling producers to earn an income for a dignified life, as well as promoting the way all prosumers would like to live. As a disruptive internet technology platform Good4Trust.org uses the power of people and of social media to strengthen the identity of the prosumers worldwide and to support socially and ecologically just producers in their endeavor to drive an economy that is like a forest, which provides better chances for us and our planet to survive.

### 5.1 Governance in Good4Trust

With every good deed the prosumers share on the platform, prosumers can receive up to 7 drops of water. When they join the platform, they are given a seed. This seed is watered by the good deeds they do, including their purchases on the platform. Every time they do a good deed or purchase, with the water they accumulate the seed slowly grows to be a seedling, then a sapling, becoming a young tree, then a mature tree, a flowering tree, and finally a fruit-bearing tree (Figure 10).

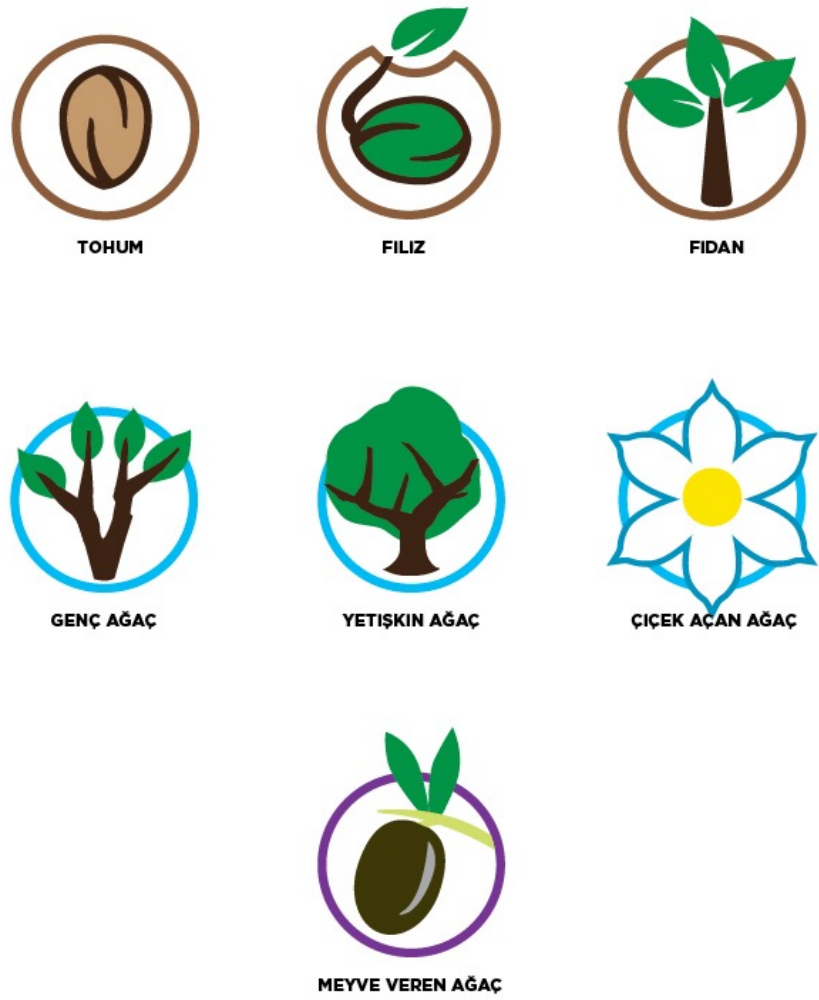

Figure 10. The 7 levels of participation through which a prosumer will move to gain governance rights from seed to seedling, sapling, to a young tree, then a mature tree, a flowering tree, and finally a fruit-bearing tree.

Through continuous use of the platform, and by fulfilling their needs from Good4Trust, they not only structure the economy into a prosumer economy, but they also gain status by growing their tree. The status gives them rights to participate in the governance of the platform. As a result the more you are engaged, the more rights you have. Good4Trust is governed by a Council of Seven, consisting of three women, three men, and one LGBTQI, who are elected among prosumers who reach the status of fruit-bearing tree. In the selection process of the Council of Seven, young trees have one vote, and mature trees have two votes. Council candidates can only be nominated by flowering trees, from the group of fruit-bearing trees. With an ordered equality election system, there is always a gender balance of 3:3:1.

The Council of Seven, primarily decides on a "consensus minus one" principle, according to which producers may join the platform based on their ecologically and socially just production systems, as well as their trustworthiness. The council does not use any criteria in line with metadisciplinary thought (Özesmi 1999), and simply because of the "problem of criterion" as it was put forward by Sextus Empiricus in "Outlines of Pyrrhonism" as a major issue in the debate between the Academic Skeptics and the Stoics. What do we know? How are we to decide in any particular case whether we have knowledge? Before we can determine what we know we must first have a method or criterion for distinguishing cases of knowledge from cases that are not knowledge. Yet, it seems that before we can determine the appropriate criterion of knowledge we must first know which particular instances are in fact knowledge. More importantly as Lyotard (1982) articulates discourses that have specific rules and criteria "is always in danger of being incorporated into the programming of the social whole as a simple tool for the optimization of its performance; this is because its desire for unitary and totalizing truth lends itself to the unitary and totalizing practice of the system's managers." As explained at the beginning The Council of Seven decides based on a shared sense of positive change in ecological and social impact. They look for difference in areas of inputs of raw materials, materials, components and all other beings such as water and energy, the production method, circularity of process and materials, work place, working conditions, wages and rights, ethical management, communication, respect and care for human and Mother Earth rights. The Council of Seven also has all the functions of a board by making policy decisions, monitoring progress of the platform, and holding the executive arm accountable. Functions of implementing sanctions, such as ending contracts with producers based on ethical breach, and resolving disputes also lies with the Council.

### 5.2 Good4Trust.org in Operation and Growing

We have established this model in Turkey and it's functioning since 2014. Since its inception in July 2016, the online bazaar has seen a considerable growth in product variety, prosumer uptake and sustained online activity (see annual reports at [https://good4trust.org/reports](https://good4trust.org/reports))

Good4Trust.org Bazaar promotes ecologically and socially just production and prosumption in order to maintain sustainability and peace. The online bazaar is a hub for dedicated producers and prosumers who base their principles on values of trust, transparency, justice and peace. The system encourages circularity, money entering the system stays in the system, and the system is growing progressively. Good4Trust.org is maintained by a 3% contribution from sales, membership from prosumers and producers who can afford. The scale of the system is constantly growing, and can be seen in full transparency at [https://good4trust.org/press](https://good4trust.org/press).

Prosumers don't consume, they purchase their needs for daily life. For example an NGO which has built a supermarket to give excess food to ones in need (a food banking system) may buy their grocery bags they provide to beneficiaries from a social enterprise producing organic cotton bags. A prosumer buys *lye* (ash-water) to wash her organic - fair-trade shirts. The ash-water producer sources her ash from the organic heirloom wheat bread baker, and so on. This is how the macro-level circular economy is built, one connection, one relation at a time until it becomes a forest.

As the forest ecosystem starts forming, also a new scholarship will start forming. We will need researchers looking at behavior of prosumers and producers, and compare that to consumers and conventional producers. The economic transactions, value chains of producers, and their supply chains and networks need to be researched, analysed, described, and quantified. We will start talking about ecolometrics rather than econometrics. We will see a new branch of science emerge Economic Ecologists, rather than Ecological Economists.

### 5.3 Becoming glocal

The expansion of the prosumer economy to different locales, cities, regions and countries is indispensable for facing our planet's global as well as local problems. Good4Trust.org is an open-source software and is given to anyone who wants to build the system with a not-for-profit social licence agreement. Currently Good4Trust.org is opening up in South Africa, and Germany in addition to Turkey, where it already exists and is developing. The challenge ahead is completing this scaling. If the prosumer economy is to succeed in contributing to the transition to an ecologically and socially just planet, we need tens of thousands of Good4Trust.org type platforms to emerge. Good4Trust then would need to attract hundreds of millions or maybe billions of prosumers, and tens of millions of producers in order to make the consumer economy obsolete. For this to happen, either one would need massive investment, or rapid replication and network effects to kick in, or both. Alternatively it becomes a movement of self-financed thousands of variants of Good4Trust type prosumer economy organisations.

As the platform and prosumer economy system proves itself stronger in different locales, replication will also follow. Currently there are many initiatives resisting or trying to cope with the existing economic system, which shows the acute need for what we are trying to establish. The prosumer economy, and Good4Trust.org specifically, could emerge as a possible future economic model globally.

### 6 Conclusion: Let's imagine and plant together!

The prosumer economy has great potential in helping to solve our planetary environmental problems. We propose a new economic governance model providing inclusive ecologically and socially just development, with the belief that it can prevent the destruction of our ecosystem and the biodiversity this planet hosts. With the Good4Trust tool, we don't preach for a prosumer economy, but put it into real action. We bring prosumers together with producers into a circular re-enforcing and eventually regenerative economy. The redistribution of wealth to the local producers is providing them with a life of dignity and a safe commercial space. The governance system is all inclusive as it is governed by the prosumers themselves. This is particularly crucial because we believe that solutions in the interest of everyone can only be achieved through grassroots governance. If we are to achieve peace and harmony with nature, then we will have to live ecologically and socially just as a society. With Good4Trust and the prosumer economy today we have taken a step towards that direction and we hope to convince more people to take steps with us, or at least create variants based on the same principles of golden rule, ecological and social justice in a producer level macro-scale, mostly local, circular economy with no leakage.

Now imagine that the whole economy of the world is a prosumer economy. A macro-scale circular economy with minimum negative or positive ecological and social impact is formed. We have an ecosystem of producers and prosumers, who have synergistic and circular relationships with deepened circular supply chains/networks. The food we buy is organic, free of toxins and healthy, we get it at a fair price. We don't think twice for our health or for our kids health. We know the people producing our food personally. We are sure that the workers behind all modes of production are socially protected, are at a legal age to work and have enough income to assure a good education for their children. Thinking globally, imagine that we have no more concerns about the extinction of animals or other living beings. We have enough fresh water for everyone. We have a pleasant and livable climate and are not scared of extreme temperatures or storms and floods that could kill us and other beings.

We have a beautiful house called the Earth that we inherited from our ancestors, now we have to do more to maintain this house for our's and the next generations' lives. For this we can be a forest, we will be a forest.

*Acknowledgements*: Neşet Kutluğ, [my team](#) and [Council of 7 Members](#) at [Good4Trust.org](#), Members of the Paradigm Shift Working Group at Greenpeace (2009-12), Michael Narberhaus, David Fell, and all [SmartCSOs](#) members, Burcu Tuncer, Gülcan Nitsch, Ümit Şahin, Ömer Madra, [Ashoka Foundation Turkey](#), Sibel Asna, Selin Gücüm, Ece Yener, Müge Nurgün, Buket Uzuner, Zeynep Kadirbeyoğlu, Gökçe Dervişoğlu, Deniz Üçok.

*Acknowledgements*: Neşet Kutluğ, my team and Council of 7 Members at Good4Trust.org, Members of the Paradigm Shift Working Group at Greenpeace (2009-12), Michael Narberhaus, David Fell, and all SmartCSOs members, Burcu Tuncer, Gülcan Nitsch, Ümit Şahin, Ömer Madra, Ashoka Foundation Turkey, Sibel Asna, Selin Gücüm, Ece Yener, Müge Nurgün, Buket Uzuner, Zeynep Kadirbeyoğlu, Gökçe Dervişoğlu, Deniz Üçok.

Chivian, E. and A. Bernstein, eds. (2008). Sustaining life: How human health depends on biodiversity. Center for Health and the Global Environment. Oxford University Press, New York.

Dittrich, P. (2012) "Rural Developement, Food Security and Nutrition" Organic Agriculture European Commission, June.

Fatheuer T, Unmüßig B, Fuhr L. (2016) Inside the Green Economy - Promises and Pitfalls. Heinrich-Böll-Stiftung / Green Books, Cambridge / Munich. p. 200 ISBN: 978-0-85784-415-6

Guevarra, L. 2010. Gen Y's green demand for the workplace. http://www.greenbiz.com/news/2010/05/19/gen-y-green-demands-workplace (accessed 4 Oct 2018).

Hesterman, O. B., Horan, D. (2017) "The demand for 'local' food is growing — here's why investors should pay attention", Business Insider, 25 Apr, 11:53 www.businessinsider.fr/us/the-demand-for-local-food-is-growing-2017-4

Hoekstra, A. Y. (2008). The water footprint of food. In J. Förare (Ed.), Water for food (p. 54). Stockholm: The Swedisch Research Council for Environment, Agricultural Sciences and Spatial Planning (Formas).

Hopwood, B., Mellor, M., O'Brien, G. (2005). Sustainable development: mapping different approaches. Sustainable Development, 13(1), 38–52. doi:10.1002/sd.244

Hutchings, J,A., and Reynolds, J.D., (2004) Marine Fish Population Collapses: Consequences for Recovery and Extinction Risk. BioScience 54:297-309 doi: 10.1641/0006-3568(2004)054[0297:MFPCCF]2.0.CO;2

IPCC (2015) "Intergovernmental Panel on Climate Change 2014". Mitigation of Climate Change, Summary for Policymakers and Technical Summary, IPCC January 2015
https://www.ipcc.ch/pdf/assessment-report/ar5/wg3/WGIIIAR5_SPM_TS_Volume.pdf

IPCC (2018) Global Warming of 1,5 °C. http://www.ipcc.ch/report/sr15/

Jones, S.E., Ho, K. Rees, C. A., Hill, J.E., Nodwell, J.R. Elliot, M.A. (2017) Streptomyces exploration is triggered by fungal interactions and volatile signals. eLife 2017;6:e21738 doi:10.7554/eLife.21738

Kanchanapibul, M., Lacka, E., Wang, X., & Chan, H. K. (2014). An empirical investigation of green purchase behaviour among the young generation. Journal of Cleaner Production, 66, 528–536. doi:10.1016/j.jclepro.2013.10.062
2018 U. Özesmi - Prosumer Economy    26/26